\documentclass{elsart}
\usepackage{graphicx}
\begin{document}
\begin{frontmatter}
\title{Chaotic vibrations in a regenerative cutting process}
\author {Grzegorz Litak}
\address{
Department of Applied Mechanics, Technical University of Lublin, 
 Nadbystrzycka~36, PL-20-618 Lublin, Poland
}
\begin{abstract}
We have analyzed vibrations generated in an orthogonal cutting process.
Using a simple one degree of freedom model of the regenerative  cutting 
we
have observed
the complex behaviour of the system. In presence of a shaped cutting
surface,
the nonlinear
interaction between
the tool and a workpiece  leads the to chatter vibrations
of periodic,
quasi-periodic or chaotic type
depending on system parameters. To describe the profile of
the surface machined by the first pass  we used a harmonic function.
We analyzed the impact phenomenon between the tool and a workpiece after
their contact loss. 
\end{abstract}

\end{frontmatter}

\section{Introduction}

The stability of a cutting process influences directly  the quality of a
final surface.  The control of process in various working conditions is an   important
problem for machining technology.
Instabilities of  process usually manifest as  harmful chatter vibrations
generated 
during the cutting.  
Recently many papers tackled that problem focusing on 
the conditions of appearing such vibrations.
Apart from regular (periodic or quasi periodic) vibrations, 
possibilities of chaotic
ones  were also
investigated theoretically
\cite{Wu85,Gra86,Gra88,Gra96,Mar88,Lit97,Wie97a,Wie97b,War00,Ste97} as
well as
experimentally
\cite{Mar88,Tan92,Pra99}. 
The aim of this paper is to explore the mechanism of
the chatter vibration generation
using a simple model of a cutting process with one degree of freedom. 
In the paper we will clarify  the role of impacts after tool-workpiece
contact loss. 

\begin{figure}[htb]
\hspace{2.5cm}
\resizebox{0.7\textwidth}{!}{%
  \includegraphics{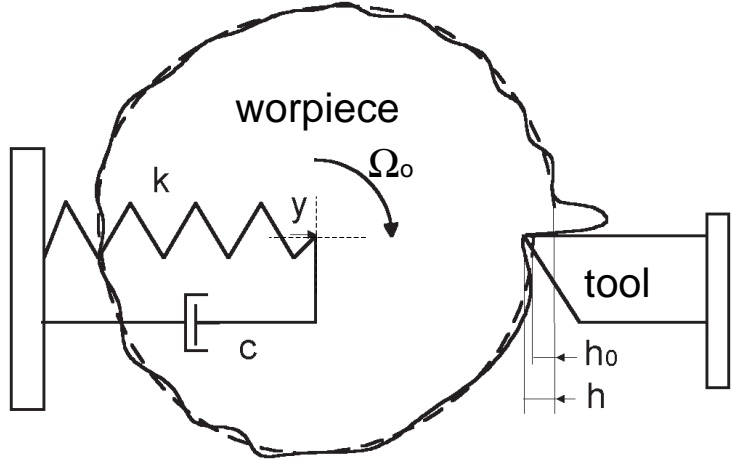}}
\vspace{0.5cm}
\caption{The model of an orthogonal turning process. The full line
illustrates 
the workpiece surface cut in the presence of chatter vibrations while
the dashed
one  corresponds to the smooth surface cut in stable conditions.}
\end{figure}

\section{The  model of a cutting process}

The one degree of freedom physical model of the orthogonal cutting process
is presented in Fig. 1.
Here $k$ denotes the effective spring of a workpiece, $c$ is the damping
coefficient, $h_0$ is the assumed  cutting depth while $h$ the
actual one, $\Omega_0$ is a rotational velocity and $v_0$ 
is a
relative
velocity between the tool and the workpiece tangent to the workpiece
surface ($v_0= 
\Omega_0r$, where $r$ is a radius of the  workpiece) . The
horizontal displacement of
the 
workpiece symmetry axis at the time $t$ is denoted by $y$.
After the first pass of 
tool the actual cutting depth $h(t)$ can be expressed as
\begin{equation}
\label{eq1}
h(t)=h_0-y(t)+y(t-T),
\end{equation}
where the $y(t-T)$ corresponds  to  the position 
of the workpiece of during the previous pass, $T$ is the
 period of revolution. In our approach, assuming constant relative
velocity 
$v_0=$ const., the actual cutting of depth  $h(t)$  will be
determined
by
the the dynamics
of the model \cite{Ste97}:
\begin{equation}
\label{eq2}
\ddot h + 2n \dot h + p^2 h = -\frac{1}{m}   {\rm sgn} (v_0+ \dot h)
F_y(h), 
\end{equation}
where $p$ is a frequency of  
free vibrations and $2n=c/m$ is a dimensionless damping coefficient.
Finally   
$F_y$ is the thrust force, a horizontal component of a nonlinear cutting
force, $m$ is the
effective mass of a workpiece.
The above equation (Eq. \ref{eq2}) can be written in terms of current
position
$y(t)$ (Eq. \ref{eq1})
corresponding to the actual position of the workpiece as
\begin{eqnarray}
\label{eq3}
\ddot y(t) + 2n \dot y(t) + p^2 y(t) &=& \frac{1}{m}  \left( {\rm sgn} (v_0- \dot
y(t))
F_y(h) \right. \\  &-&  \left.{\rm sgn} (v_0- \dot
y(t))
F_y(h_0)\right). \nonumber 
\end{eqnarray}
 
The thrust force $F_y$ is  mainly based  on a  dry friction part
(between the tool and a chip) with
power law dependence on the actual cutting
depth $h$ \cite{Ste97}: 
\begin{equation}
\label{eq4}
F_y(h)= \Theta(h) K w(h)^{3/4},  
\end{equation}
where $K$ denotes the cutting resistance, $w$ is a chip width 
and $\Theta$ is the Heaviside step function. 
To go further in our analysis we have assumed, in the first
approximation,  that  $y(t-T)$
can be
described by a periodic function
\begin{equation}
\label{eq5}
y(t-T) = a \cos(\omega t-\phi) = a \cos(\frac{\omega x}{v_0}-\phi),
\end{equation}
where $\omega$, $a$ and $\phi$ are the period, amplitude and phase of a
surface shape
modulation, respectively.
$x(t)$
denotes the
relative distance passed by the tool on the cylindrical surface of
a rotating workpiece.
\begin{figure}[htb]
\resizebox{0.8 \textwidth}{!}{%
 \includegraphics{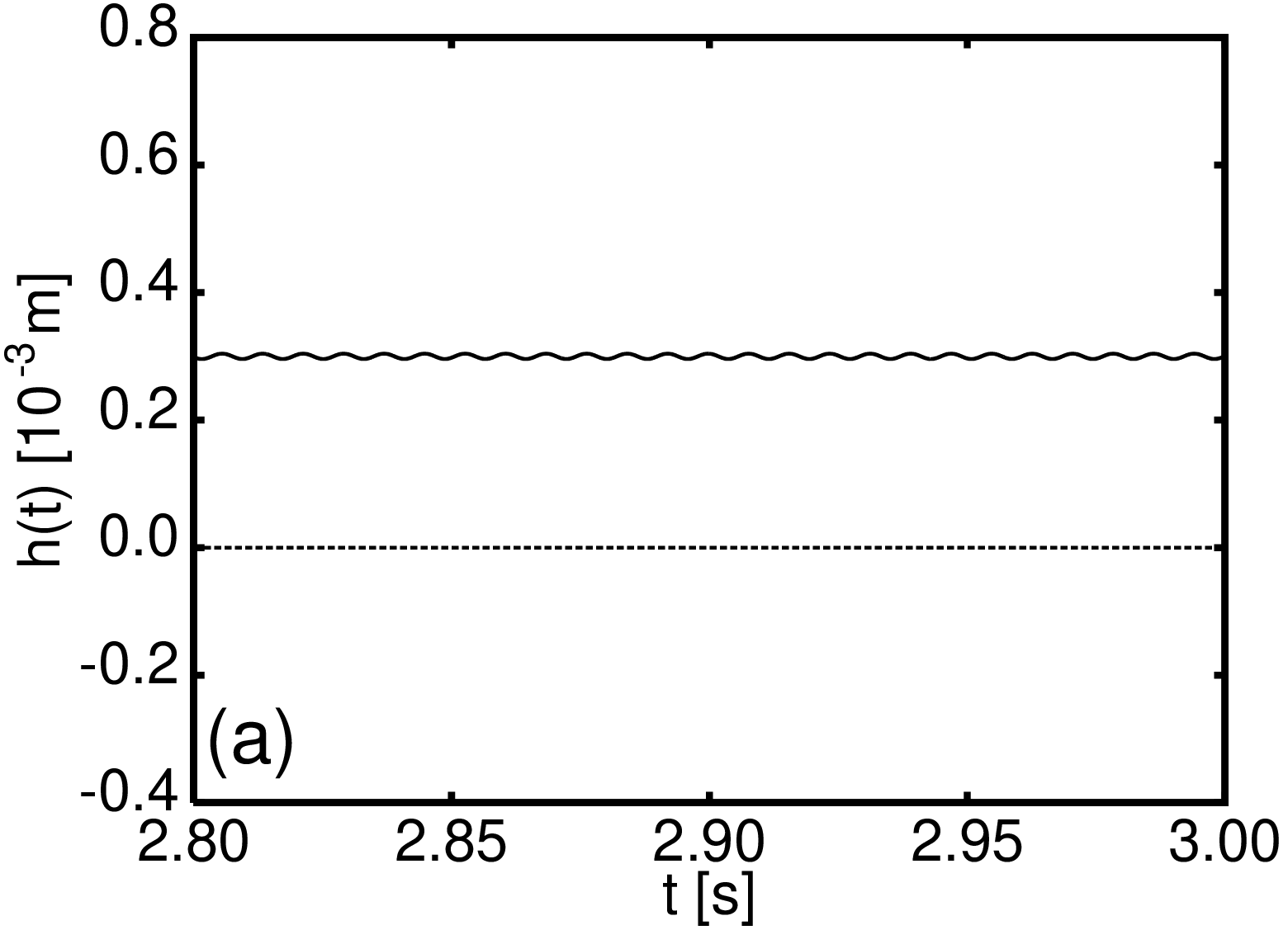}
\hspace{8cm}
\includegraphics{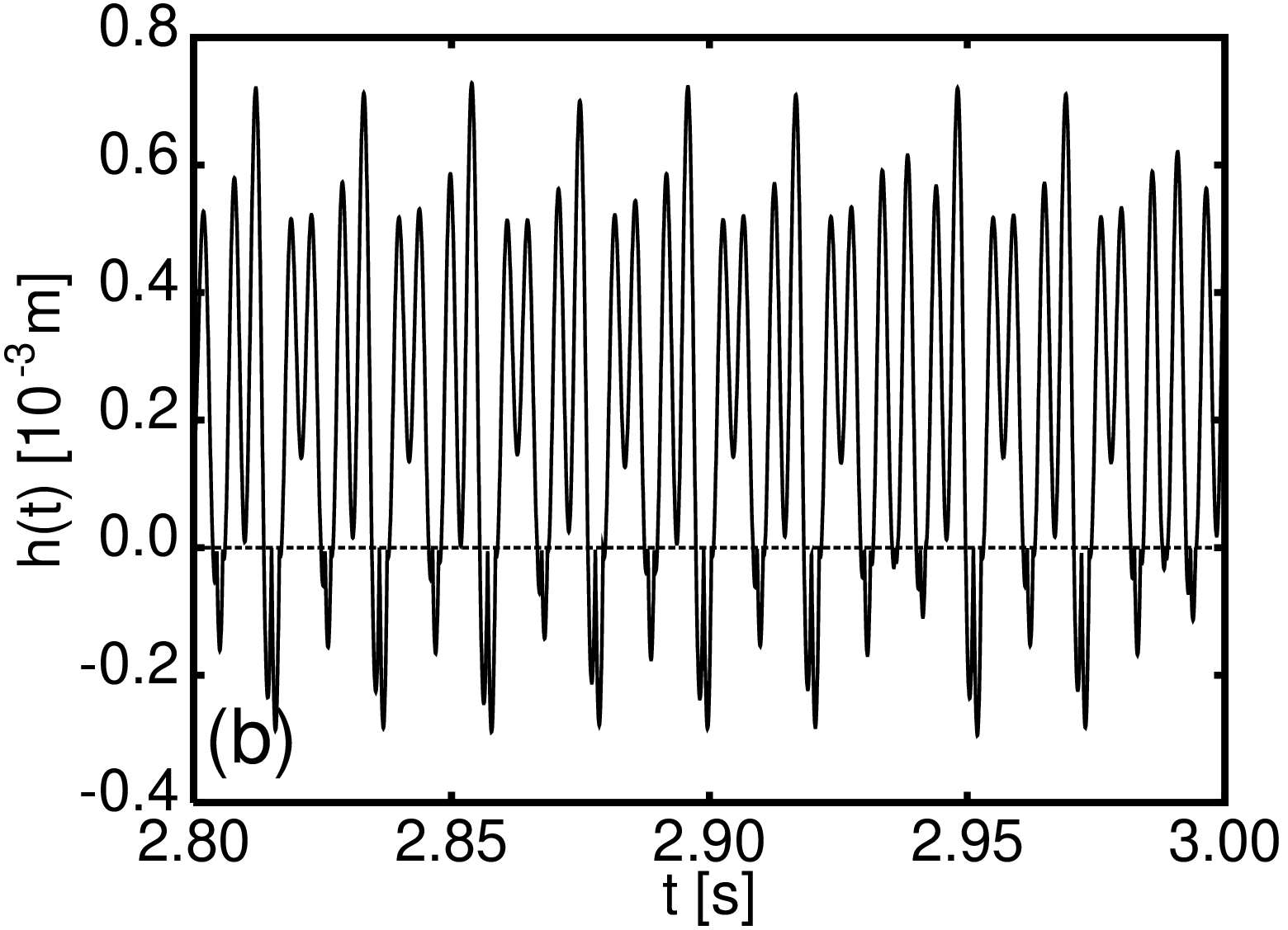}}

\vspace{-1cm}
\resizebox{0.8 \textwidth}{!}{%
\includegraphics{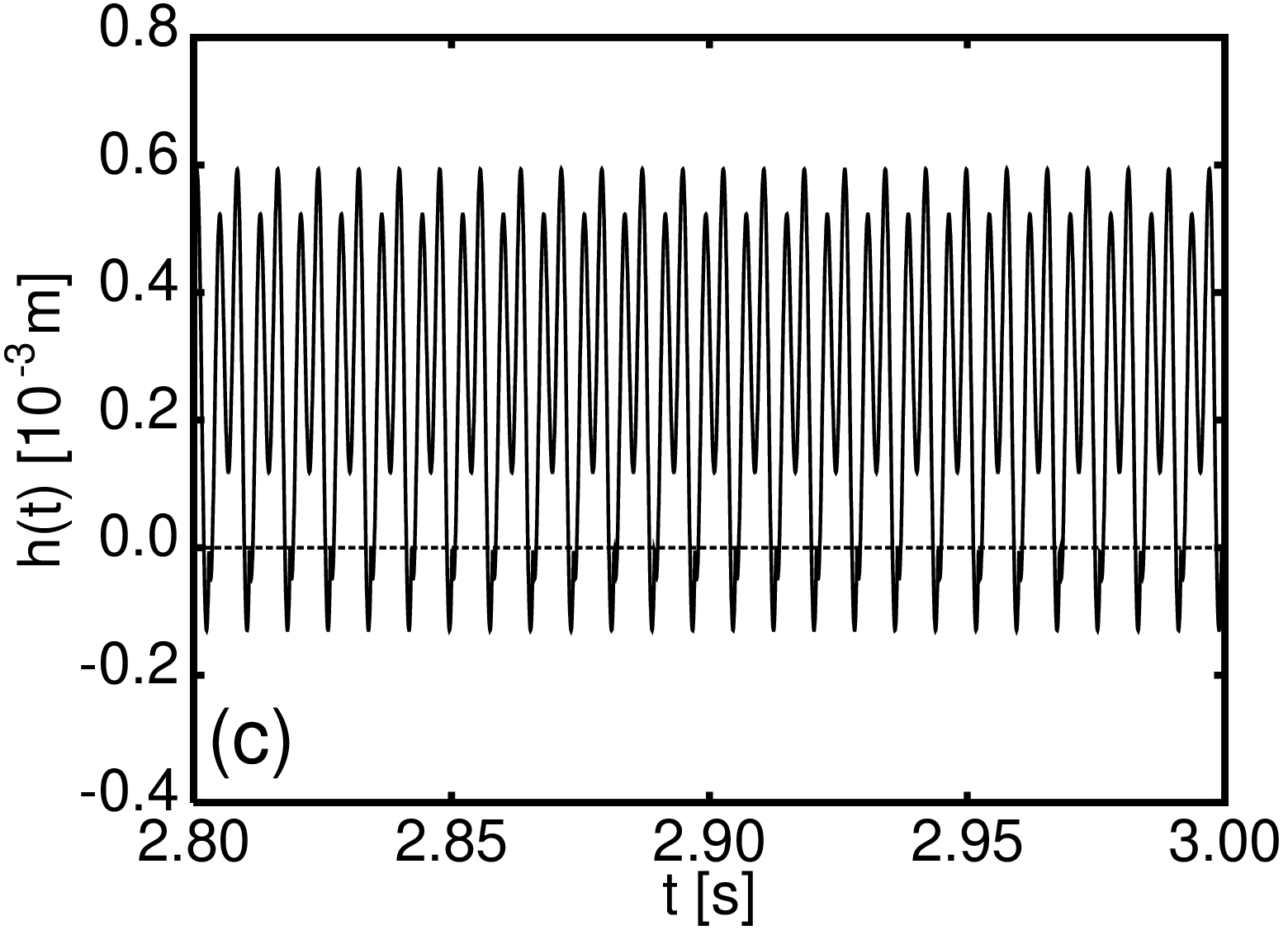}
\hspace{8cm}
\includegraphics{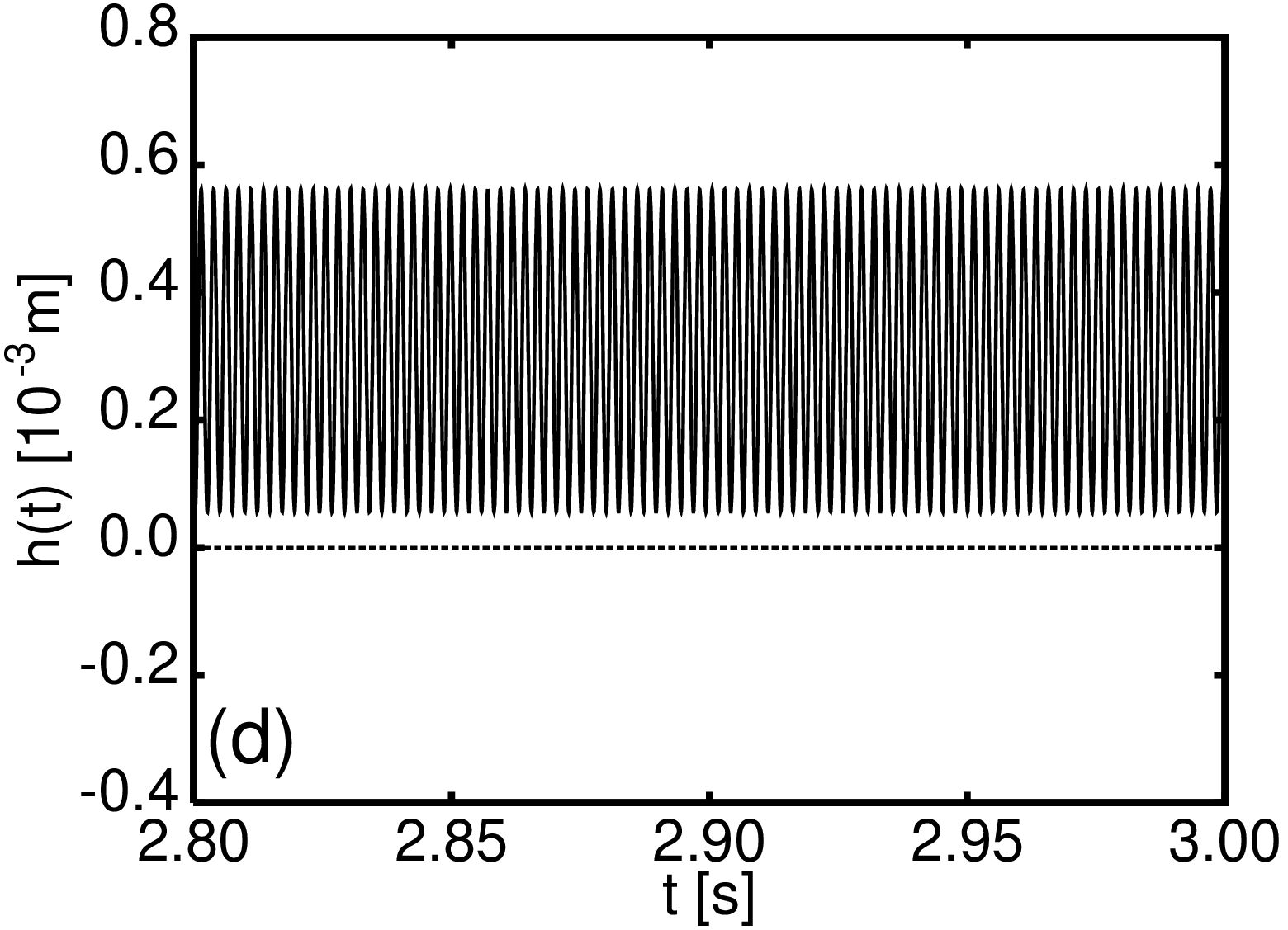}
}
\caption{Time histories of workpiece motion $y(t)$ for various frequency
$\omega$: (a) $\omega=800$ rad/s,
(b) $\omega=1200$ rad/s, (c) $\omega=1600$ rad/s, (d) $\omega=2600$ rad/s}
\end{figure}
Note that our model (Fig. 1) includes also the contact 
loss between the tool and a workpiece. 
Such phenomenon can be particularly important during the  cutting
process with a high speed.
Here we assume the restitution parameter $\beta \le  1$ 
connected with the impact after contact loss.
\begin{equation}
\dot y(t^+) = -\beta~ \dot y(t^-), 
\end{equation}
where $t^+$ and $t^-$ denotes 
time before and after impacts, respectively.
The above assumptions  (Eqs. 5,6)
will be crucial in our investigations in the next section 
where we discuss the results of simulations.

\section{Numerical simulations}

Equation 2 has been solved numerically for the realistic
system parameters  \cite{Mar88,Lit97,War00}: $K=1.25 \times 10^9 {\rm~
N/m}^2$,
$w=3.0 \times
10^{-3}{\rm~ m}$,
$h_0=0.3 \times 10^{-3} {\rm~ m} $,
 $p=816 {\rm rad/s}$, $m=17.2 {\rm kg}$, $a=0.2\times 10^{-3} {\rm~ m}$
$\phi=\pi/2$, 
 $\beta=0.75$ and a relatively small damping
$n=4.3
{\rm~s}^{-1}$.
In Figs. 2a-d we show time histories of cutting depth $h(t)$ for
various
frequency. The negative value of $h$ means that the tool lost a
contact with the workpiece. 
Fig. 2a presents  the results for  cutting with 'parametric excitation'
(Eq. 5)  $\omega=800$ rad/s. 
Note that,  this case corresponds to a relatively small velocity
$v_0$ (but still $v_0 > \dot y$).
One can see that the cutting process is very smooth with small
fluctuations
around assumed $h_0$. In contrary to that Fig. 2b (for a larger
frequency $\omega=1200$ rad/s) 
shows unstable cutting with large fluctuation of a cutting depth $h$. The
time history
$h(t)$
manifests the contact loss phenomenon. Moreover, vibrations generated
in the system look aperiodic. 
It could be caused by the lack of synchronisation between impacts
incidents and 
the driving frequency $\omega$. 
Thus the aperiodic part
is appearing after the escape of tool  into the region of negative $h <
0$ (Fig. 2b).
On the other hand the same system but  $h > 0$ seems to be more regular.
      Such behaviour is the benchmark of an intermittent mechanism of chaotic
motion \cite{Pom80}. Surely, as in other mechanical examples \cite{Cha96}
appears due to impacts.
Increasing the  $\omega$  ($\omega=1600$ rad/s in Fig. 2c) we transit back
to 
regular vibrations of cutting depth $h(t)$. Here, in spite of  a large
amplitude and a tool--workpiece contact losses the impacts are
synchronised with a 'parametric forcing' $\omega$.
Finally, for a relatively large $\omega$  ($\omega=2600$ rad/s in Fig. 2c)
which corresponds to cutting with a high speed $v_0$, the vibration
amplitude slightly decreases. 

\begin{figure}[htb]
\resizebox{0.8 \textwidth}{!}{%
\includegraphics{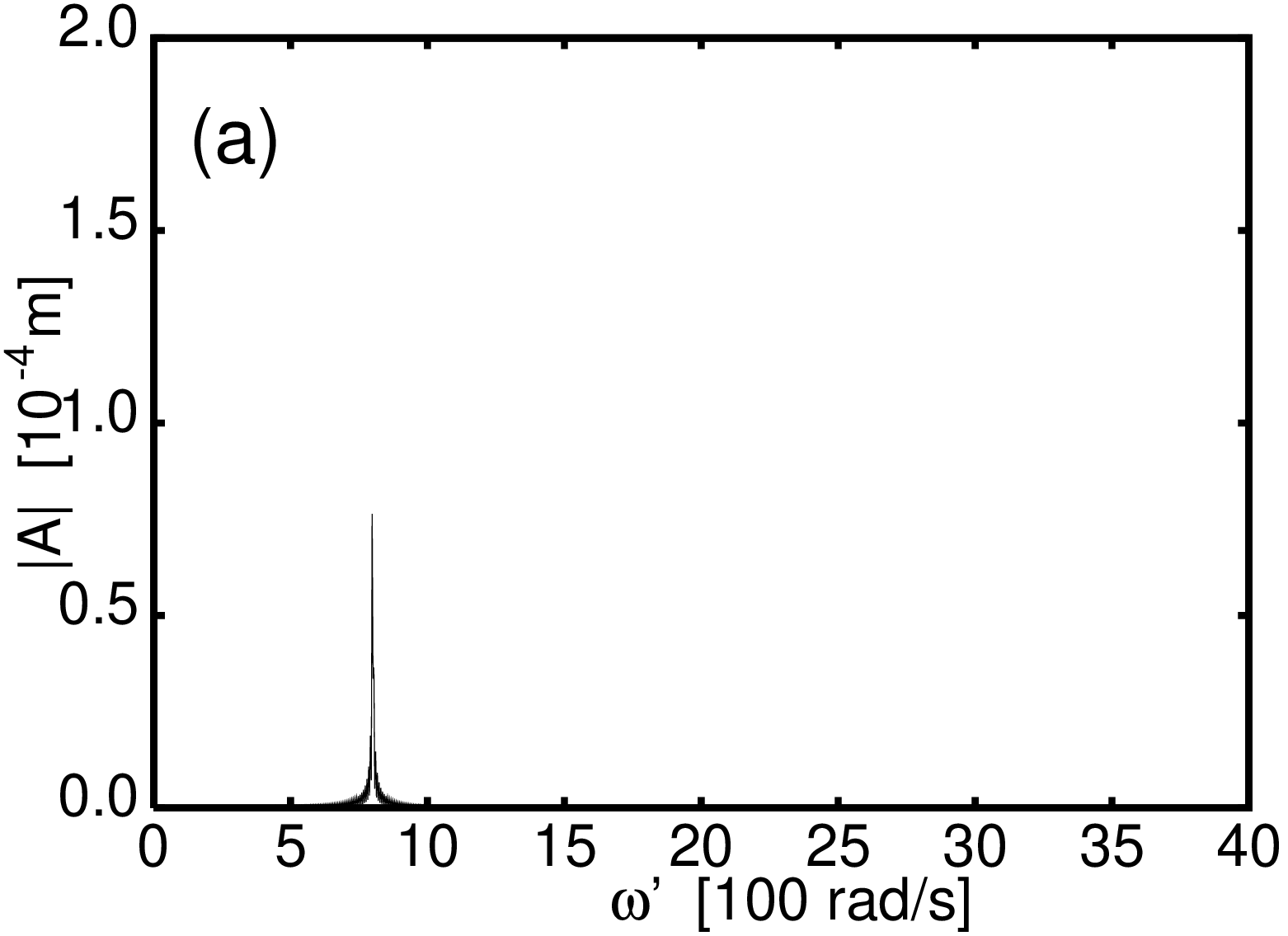}
\hspace{9.5cm}
\includegraphics{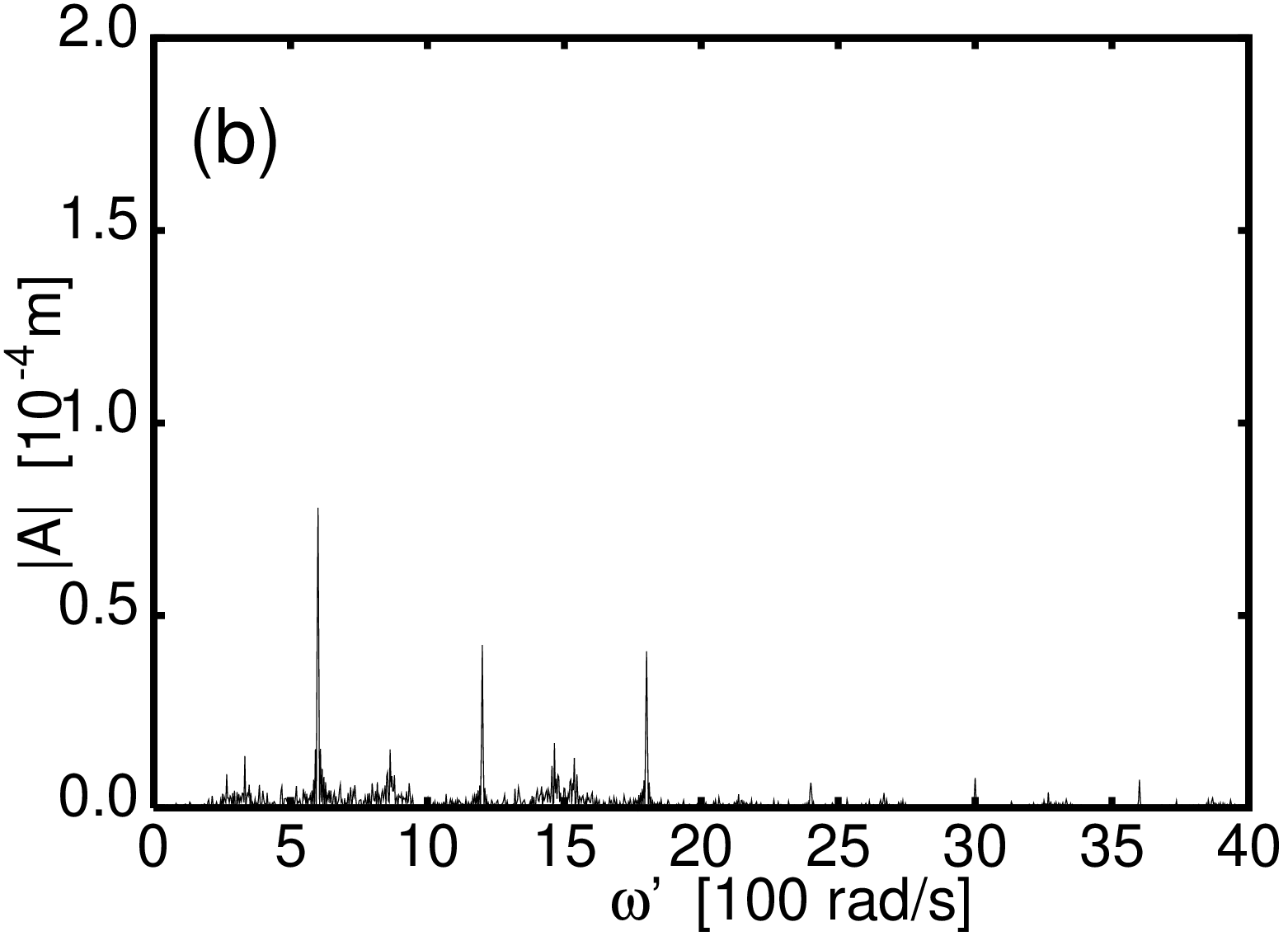}}

\vspace{-1cm}
\resizebox{0.8 \textwidth}{!}{%
\includegraphics{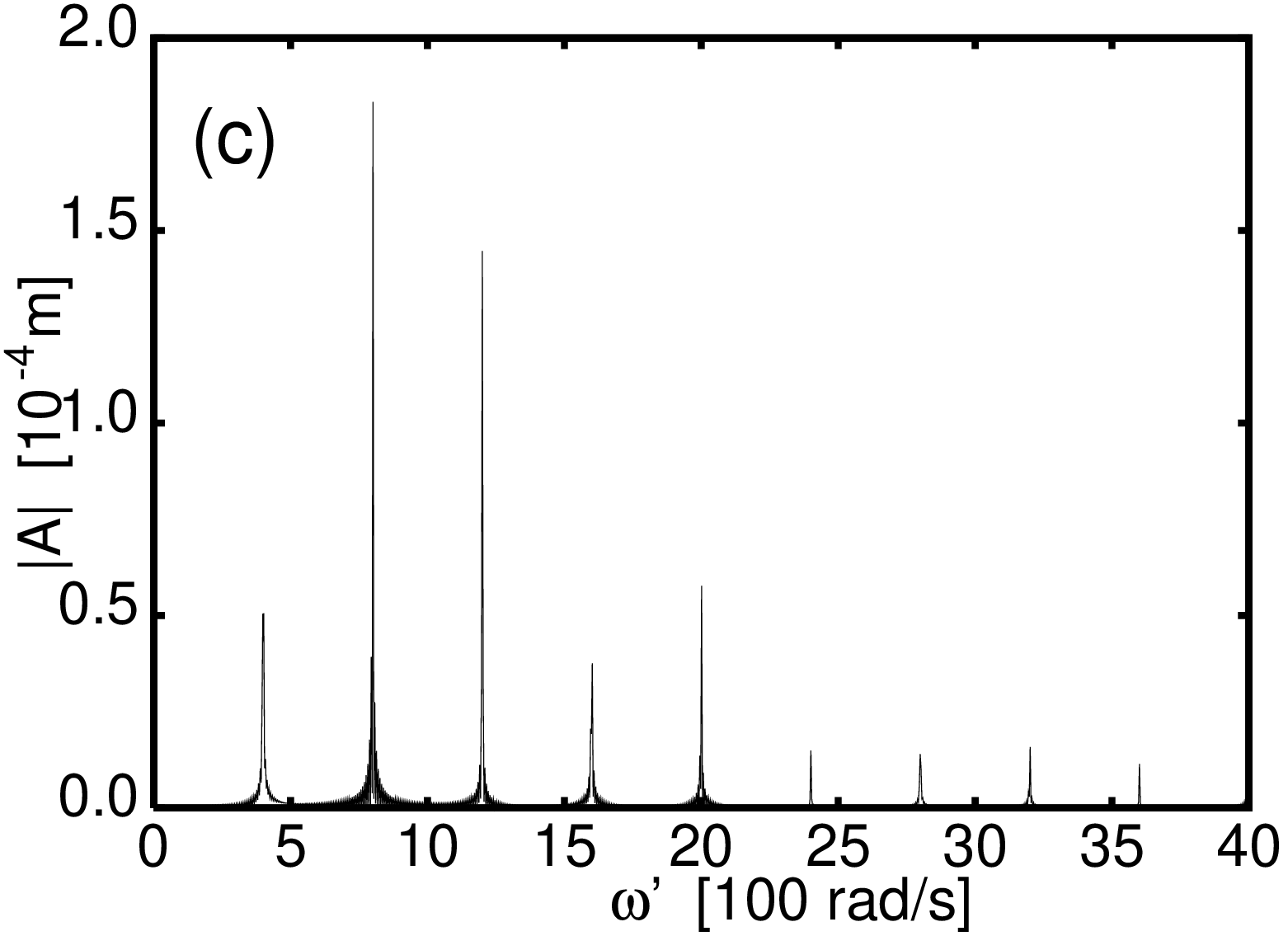}
\hspace{9.5cm}
\includegraphics{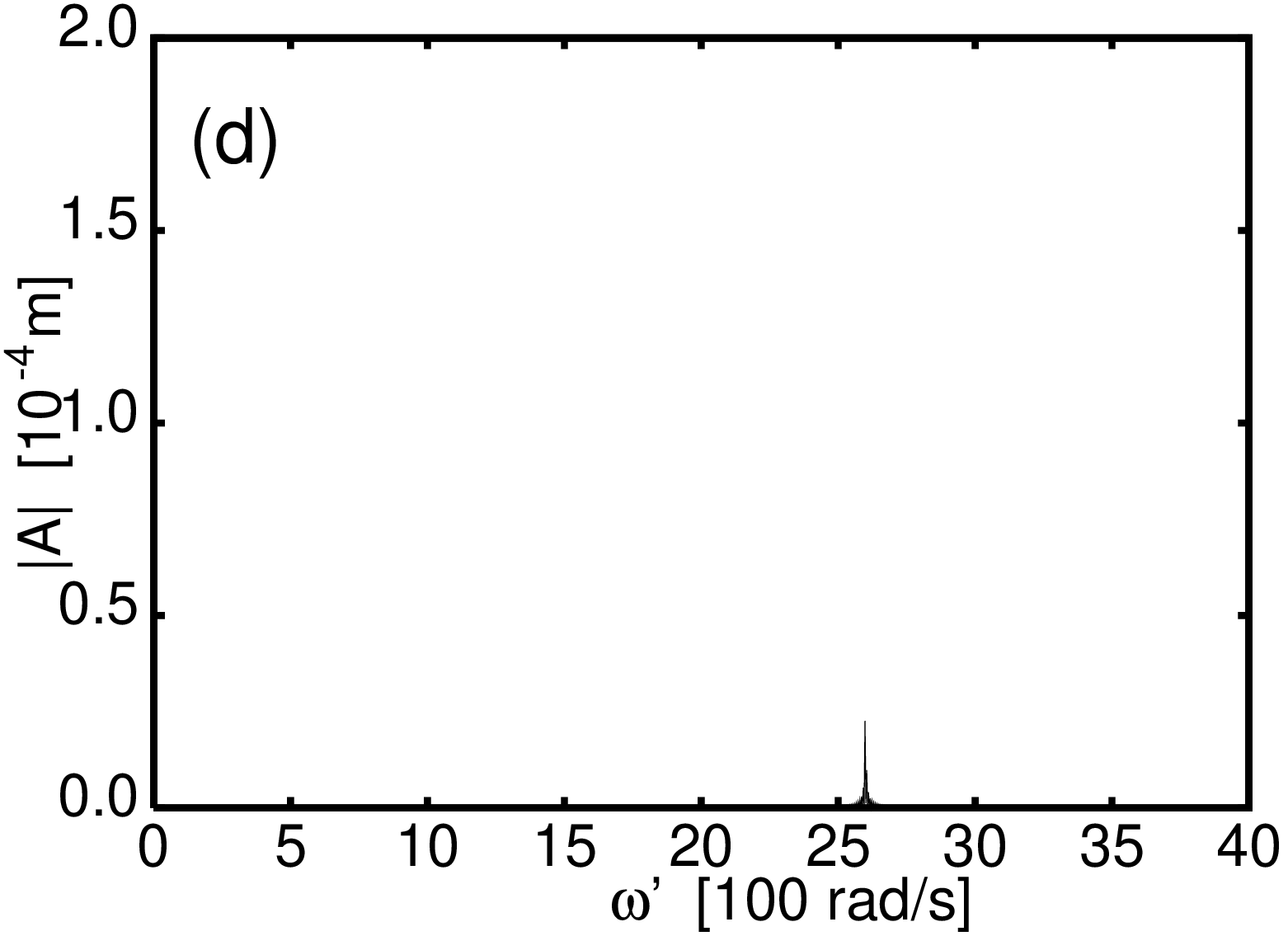}
}
\caption{The amplitude of workpiece  vibrations  $y$ versus frequency
$\omega'$, The system parameters have been chosen as in Fig. 2.}
\end{figure}

To explore the vibrations of the system we have done the Fourier transform
for each of above cases. 
Figs. 3a--d present the amplitude of workpiece  vibrations  $y(t)$ versus
frequency $\omega'$ with
the system parameters as  in Fig. 2.
Figs. 3a and 3d show spectra with a single frequency corresponding to 
driving one $\omega$ indicating on synchronized motion of a workpiece.
Fig. 3b has a combined spectrum of singular frequencies and continues
interval of  $\omega'$. This is a typical output of  chaotic systems.  
In  Fig. 3c one can find a lot of characteristic frequencies 
$\omega'$ which are  represented by natural  magnification
 of $\omega/4$ ($4\omega'/\omega= l$, where $l$ is a natural number).

Interestingly, for an small driving frequency $\omega=800$ rad/s,
which
corresponds to a small rotational
velocity of a workpiece, fluctuations of cutting depth $h$  
are small
Fig. 2a. Physically, in this case,  the workpiece in its motion $y$ can
follow
the
driving
'parametric forcing' by the initial shape modulation  $y(t-T)$.  
This is resulting
in relatively large vibration amplitude  of the workpiece (Fig. 3a).
For a  large  driving frequency  $\omega=2600$ rad/s we obtain
the opposite situation. Now fluctuations of  $h$ have larger amplitude
(Fig. 2d) while vibrations $y$  smaller one (Fig. 3d). In this case
 a workpiece, due to  its large inerta, can not follow the changes
of the initial shape forcing the large variations of $h$ via Eq. 1.

\section{Summary and  conclusions}

We have  considered  vibrations of a tool-workpiece  system in a
orthogonal turning
process. The simple one degree of freedom model we used includes the
basic phenomena as 
friction between a
chip and the tool, nonlinear - power low character of the
cutting force
expression as well as the possibility of a contact loss between
the  tool and  the workpiece.   
In our model
we observe
the complex behaviour of the system. In presence of a shaped   
initially cut
surface,
the nonlinear
interaction between
the tool and a workpiece  leads the to chatter vibrations
of a periodic,
quasi-periodic or chaotic type. The main role in chaotic motion is played
by impacts generated after the a tool-workpiece contact losses.
Clearly, they enable an intermittent transition from  a regular
to chaotic  system behaviour.   
In spite of the fact that our results are obtained by very simple model
we believe that we found the important mechanism of cutting instabilities 
by an  
impact phenomenon.

\begin{ack}
The work has been partially supported by Polish State Committee for
Scientific
Research (KBN) 
under the grant No. 126/E-361/SPUB/COST/T-7/DZ 42/99.
\end{ack}

\end{document}